\documentclass{article}

\usepackage{PRIMEarxiv}

\usepackage[utf8]{inputenc} % allow utf-8 input
\usepackage[T1]{fontenc}    % use 8-bit T1 fonts
\PassOptionsToPackage{hyphens}{url}
\usepackage{hyperref}       % hyperlinks
\usepackage{booktabs}       % professional-quality tables
\usepackage{amsfonts}       % blackboard math symbols
\usepackage{nicefrac}       % compact symbols for 1/2, etc.
\usepackage{microtype}      % microtypography
\usepackage{lipsum}
\usepackage{fancyhdr}       % header
\usepackage{graphicx}       % graphics
\graphicspath{{media/}}     % organize your images and other figures under media/ folder

\usepackage{tabularx}
\usepackage{booktabs}
\usepackage{ragged2e}

\title{QB4AIRA: A Question Bank for AI Risk Assessment
}

\author{
  Sung Une Lee, Harsha Perera, Boming Xia, Yue Liu, Qinghua Lu, Liming Zhu, Olivier Salvado, Jon Whittle \\
  Data61, CSIRO, Australia\\
  \texttt{firstname.lastname@data61.csiro.au}
}

\begin{document}
\maketitle

\begin{abstract}
The rapid advancement of Artificial Intelligence (AI), represented by ChatGPT, has raised concerns about responsible AI development and utilization. Existing frameworks lack a comprehensive synthesis of AI risk assessment questions. To address this, we introduce QB4AIRA, a novel question bank developed by refining questions from five globally recognized AI risk frameworks, categorized according to Australia's AI ethics principles. QB4AIRA comprises 293 prioritized questions covering a wide range of AI risk areas, facilitating effective risk assessment. It serves as a valuable resource for stakeholders in assessing and managing AI risks, while paving the way for new risk frameworks and guidelines. By promoting responsible AI practices, QB4AIRA contributes to responsible AI deployment, mitigating potential risks and harms. 
\end{abstract}

% keywords can be removed

\section{Introduction}
Artificial Intelligence (AI) has witnessed remarkable advancements in recent years, permeating numerous domains and transforming various aspects of our lives. 
However, this rapid advancement of AI technologies has also raised concerns about their responsible development and application.
For instance, while large language models, such as ChatGPT, have demonstrated impressive capabilities in generating text-based responses across a wide range of queries, they exhibit limitations and risks such as hallucination, generating harmful content, and user's overreliance \cite{OpenAI2023}.
Incidents involving AI systems, documented in repositories like the AI Incident Database\footnote{\url{https://incidentdatabase.ai/}}, also highlight the urgency of addressing AI risks. As of July 2023, this repository has cataloged over 2,800 unfavorable incidents caused by AI systems worldwide, emphasizing the need for comprehensive risk management.

To ensure the responsible development, deployment, and utilization of AI systems, the concept of Responsible AI (RAI) has emerged. RAI involves the practice of developing AI systems that benefit individuals, groups, and society as a whole, while minimizing the risk of adverse consequences~\cite{lu2022responsible, xia2023systematic}. Risks in this context encompass the likelihood and consequences that arise when the development and use of AI systems fail to align with RAI principles, such as fairness, privacy, explainability, and accountability throughout the AI development process. Therefore, it is crucial to ensure that the development and use of AI systems adhere to these RAI principles.

% Responsible AI (RAI) refers to the practice of developing, deploying, and utilizing AI systems in a manner that benefits individuals, groups, and society as a whole, while minimizing the risk of adverse consequences~\cite{lu2022responsible, xia2023systematic}. The term \textit{risks} typically encompasses the likelihood and consequences that arise when the development and use of AI systems fail to align with RAI principles, such as fairness, privacy, explainability, and accountability throughout the AI development process. Therefore, ensuring that the development and use of an AI system adhere to these RAI principles is crucial.

Several organizations have developed principle-driven RAI risk assessment frameworks to guide the implementation of RAI principles. Examples of these frameworks include the US NIST AI Risk Management Framework\footnote{\url{https://www.nist.gov/itl/ai-risk-management-framework}} and the EU Assessment List for Trust AI Framework\footnote{\url{https://futurium.ec.europa.eu/en/european-ai-alliance/pages/welcome-altai-portal}}. 
While these frameworks provide valuable guidance for evaluating and mitigating potential risks associated with AI systems, 
%there is a lack of a comprehensive and organized synthesis of dedicated AI risk assessment question banks that are \textit{comprehensive, connected, and layered} in nature. 
there is a lack of a comprehensive and well-structured AI risk assessment question bank that incorporates interconnected and layered questions and connectivity to other standards and frameworks \cite{xia2023systematic}.
Existing question banks often focus on specific aspects, overlooking the holistic nature of RAI principles. For instance, Liao et al. \cite{liao2020questioning, liao2021question} introduced a question bank specifically focused on transparency and explainability, neglecting other crucial principles. Additionally, the US Department of Energy's AI Risk Management Playbook\footnote{\url{https://www.energy.gov/ai/doe-ai-risk-management-playbook-airmp}} provides extensive guidance on identifying AI risks and recommending mitigations, but it lacks integration with diverse standards, frameworks, and related resources.

To bridge this gap, we propose QB4AIRA as a comprehensive and connected resource for AI risk assessment. 
%The overall research steps are depicted in Figure~\ref{fig:methodology}. 
In this paper, we present the development of QB4AIRA by conducting a review of existing frameworks identified in \cite{xia2023systematic}, analyzing their core characteristics, and evaluating their alignment with RAI principles.
We selected five frameworks to develop the question bank based on (i) global recognition, (ii) inclusion of risk assessment questions, and (iii) representation of different regions and industry leaders.
These frameworks encompass various geopolitical contexts, including the \textit{EU Trustworthy AI Assessment List, Canada Algorithmic Impact Assessment\footnote{\url{https://www.canada.ca/en/government/system/digital-government/digital-government-innovations/responsible-use-ai/algorithmic-impact-assessment.html}}, Australia NSW AI Assurance Framework\footnote{\url{https://www.digital.nsw.gov.au/policy/artificial-intelligence/nsw-artificial-intelligence-assurance-framework}}, Microsoft (MS) Responsible AI Impact Assessment Guide\footnote{\url{https://blogs.microsoft.com/wp-content/uploads/prod/sites/5/2022/06/Microsoft-RAI-Impact-Assessment-Guide.pdf}}, and the US NIST AI Risk Management Framework.}

% From these frameworks, we collected a total of 382 questions: 126 from the EU framework, 66 from Canada, 82 from NSW, 78 from MS, and 30 from NIST. Through a process of refinement involving rephrasing and merging duplicate questions, we condensed the collection into 293 questions. To ensure systematic categorization, we classified the questions based on the eight core concepts of Australia's AI ethics principles \cite{aiethicsaus2019}. Furthermore, we conducted a fine-grained concept mapping, inspired by Trochim's approach \cite{trochim1989introduction}, to identify common and crucial risk themes. 
Through iterative refinement, we condensed 382 questions into 293, systematically categorized them based on Australia's AI ethics principles \cite{aiethicsaus2019}, and conducted concept mapping to identify common risk themes \cite{trochim1989introduction}. 
The hierarchy of risk questions was structured using a decision tree \cite{song2015decision}, considering their level and sequence. 
QB4AIRA has undergone evaluation through two case studies.

%The rest of the paper discusses the design approach, development challenges, architecture, key features and the uses of QB4AIRA.  

% background & motivation
% current problems
% justification
%  - why do we need a new question bank? 
%    . As a fundamental knowledge base, it needs to effectively respond to diverse stakeholders, meet different requirements, and accommodate varying levels of detail. 
%    . Yet, the previous studies lack in..., no framework can cover all AI ethics principles.... 
%    
% aims
% research questions
% contributions
% structure of the paper

%\section{Literature review}
% background information
% related work

%\section{Methodology}
% data collection
% table/figure- risk questions collected from the five frameworks
% data analysis
% 

\section{The Question Bank: QB4AIRA}
% - show the simple structure of the question bank (high-level architecture)
% - show example of the questions (and use appendix to show full question bank?)

% \subsection{Overview}
%This section introduces QB4AIRA, a comprehensive risk assessment tool for RAI that is designed to meet the requirements of diverse users, including individuals at the C-level, project manager level, developer level, and beyond.
QB4AIRA is a collection of risk questions aimed at supporting the ethical development of AI systems.
It is designed to meet the requirements of diverse users, including individuals at the C-level, project manager level, developer level, and beyond.
%The necessity for such a question bank is evident, given the complex and multifaceted nature of AI risks that can arise from various sources, including biased data, algorithmic errors, unintended consequences, and more. 
%To effectively manage the complex and multifaceted nature of AI risks, organizations require a structured and systematic approach that encompasses all relevant areas of risk management. 
%The purpose of QB4AIRA is to address the unique challenges and considerations associated with AI risk assessment while providing a comprehensive and effective tool for this purpose. 
QB4AIRA also tackles the distinct challenges and considerations linked to AI risk assessment and helps identify areas where risk management practices may be lacking.

\subsection{Architecture}
\begin{figure*}[htb]
  \centering
  \includegraphics[width=\textwidth]{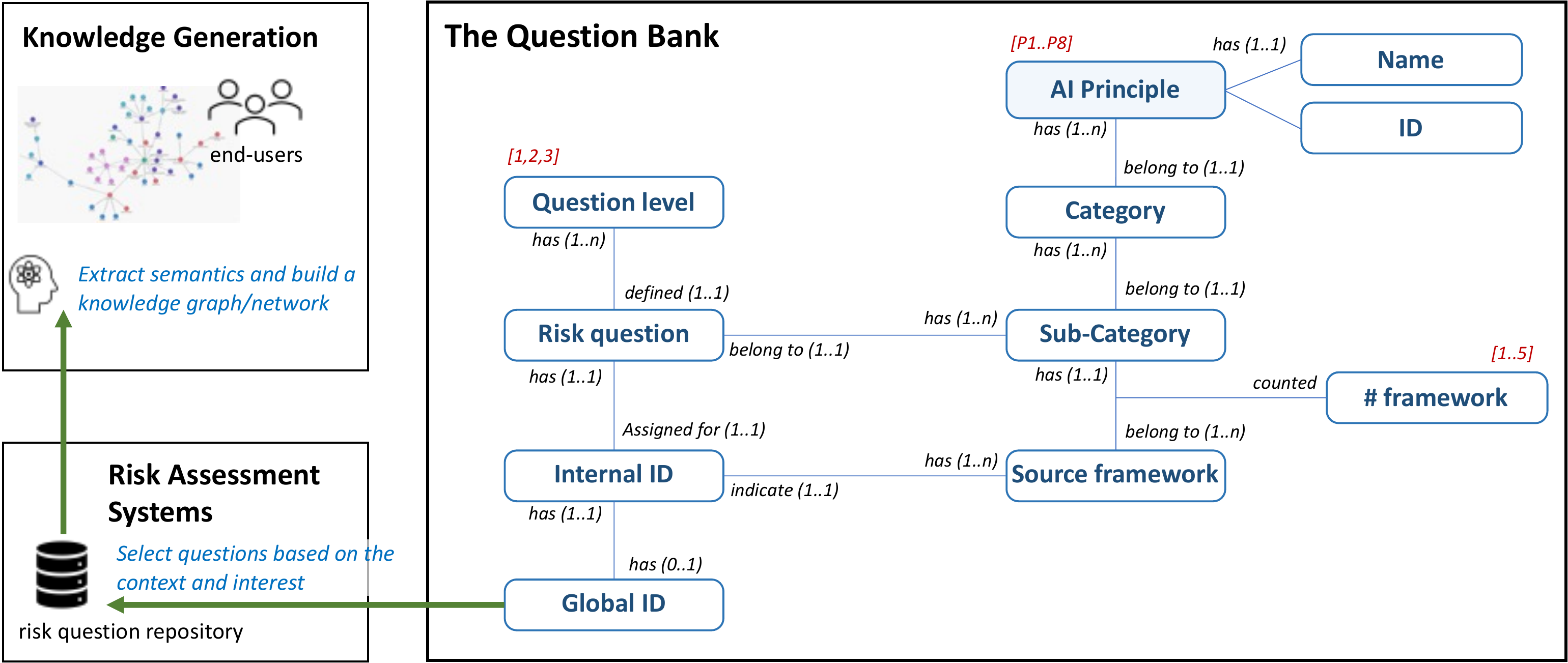}
  \caption{The architecture of the question bank.}
  \label{fig:QB_Architecture}
\end{figure*}

QB4AIRA is structured to ensure comprehensive coverage of key components and features a clear focus on RAI (Figure~\ref{fig:QB_Architecture}). 

\textbf{AI Ethics principle:} We have chosen Australia's ethical AI principles as they closely align with similar principles \cite{jobin2019global, fjeld2020principled} from around the world.
These principles serve as the overarching categories for the question bank including \textit{Human, societal and environmental wellbeing, Human-centered Value, Fairness, Privacy and Security, Reliability and Safety, Transparency and Explainability, Contestability,} and \textit{Accountability}. 
%Each principle is composed of key elements such as risk categories and questions.
To facilitate navigation and identification, a two-digit sequential identifier is assigned to each principle (P1 to P8).

\textbf{Categories:} Each principle can be further divided into multiple categories, each representing a specific aspect of the principle. Each \textit{Category} is exclusively associated with one principle.
The categories act as common risk themes that are clear, specific, and relevant to the AI ethics principle they represent. Additionally, a category may contain one or more \textit{Sub-Categories}, providing even more granularity and specificity to the questions and facilitating a better understanding of the different dimensions of each category.

% need to rewrite the following section
\textbf{Risk question:} This component consists of a set of questions associated with each category (examples in Table \ref{tab:concept}). 
These questions are to identify potential risks during AI system development, ensuring comprehensive scrutiny and resolution of all potential risks.
%For example, the question \textit{"Do you establish mechanisms that facilitate the system’s auditability?"} is to assess whether appropriate mechanisms are in place to enable auditing and review of the AI system's decision-making process. 
%An unauditable system may hinder the identification of errors or biases, leading to negative consequences for stakeholders and society as a whole.

\textbf{Question ID:} Question IDs are used to uniquely identify each question within the question bank. 
\textit{Internal IDs} are employed for managing the questions internally, including tracking the source framework of each question. 
This metadata is crucial for maintaining mapping transparency and accuracy. 
% It is important to include the original source of each question as a part of its metadata or associated information.
As a result, all original questions (a total of 382) have been assigned \textit{Internal IDs}.
%It enables users to determine whether it is relevant to their specific needs or situation. 
Additionally, the refined questions (293 in total) have been assigned \textit{Global IDs}, which is a universal reference number for external entities. 
%These IDs can be utilized by risk assessment systems and users who want to reference specific questions within QB4AIRA.

\textbf{Question level:} Each question has a question level (1, 2, or 3), indicating its importance and level of detail within the risk category. 
The question level is essential for facilitating tiered risk assessment (discussed further in Section \ref{sec:tiered}). 

\textbf{Source framework:} It provides a link to the original framework of each question. 
This helps to ensure transparency and accuracy as it allows users to easily access and verify the original source of the questions.
% The number of frameworks referencing each sub-category varies from one to five, indicating the level of consensus and coverage among the selected frameworks regarding specific sub-categories.
We indicate the extent to which each sub-category is addressed as a common concept across frameworks by including the number of source frameworks, ranging from 1 to 5.
%This information provides insights into the level of consensus and coverage among the selected frameworks regarding specific sub-categories.

\textbf{External entities:} QB4AIRA can be utilized by risk assessment systems or individual users. 
They can access QB4AIRA via \textit{Global IDs} and select the most appropriate ones based on their specific use case, context, and interests.
By extracting the relevant concepts and relationships from the selected questions, a knowledge graph/network can be built to provide more personalized and accurate responses to user queries.

\subsection{Key features}

\subsubsection{Concept map}
We employed a concept mapping approach \cite{trochim1989introduction} to analyze and group related risk questions, thereby identifying AI risk themes.
This process clarifies the relationships between different risk themes and provides a comprehensive overview of the key areas of concern for AI systems. 
%The resulting concept map can be used as a tool for communication, collaboration, and decision-making among stakeholders involved in AI risk assessment.

Table \ref{tab:concept} provides an overview of the identified risk concepts, including the number of categories (themes) for each AI principle, the total number of questions, and other pertinent information.
This facilitates a better understanding of the distribution of questions across different principles and themes.
With a total of 293 questions, QB4AIRA covers 31 risk themes and 68 sub-themes, demonstrating its comprehensive coverage of various aspects of RAI.

\begin{table}[ht]
  \caption{The number of common risk themes (category and sub-category) and questions categorized by AI ethics principles; there are also selected examples of the risk themes and questions for \textit{Accountability, Transparency, and explainability,} and \textit{Reliability and safety.}}
  \label{tab:concept}
  \scriptsize
  \begin{tabularx}{\linewidth}{p{0.15\linewidth}p{0.09\linewidth}p{0.12\linewidth}p{0.4\linewidth}X}
    \toprule
    Principle & \#Category & \#Sub-Category & \#Question & \#Framework \\
    \midrule
    Human wellbeing & 5 & 12 & 71 & 5 \\
    Human values & 3 & 5 & 19 & 4 \\
    Fairness & 3 & 5 & 38 & 5 \\
    Privacy/security & 4 & 7 & 46 & 5 \\
    Reliability/safety & 5 & 15 & 49 & 4 \\
    Transparency/explainability & 4 & 7 & 30 & 5 \\
    Contestability & 2 & 4 & 4 & 2 \\
    Accountability & 5 & 13 & 36 & 5 \\
    \midrule
    & 31 & 68 & 293 & 2-5 \\
    \midrule
    \multicolumn{5}{l}{\textit{(Examples of risk themes and questions)}} \\
    Accountability & auditability & traceability/logging & .Do you establish mechanisms that facilitate the system’s auditability, such as ensuring traceability and logging of the AI system’s processes and outcomes? & EU\\
    & & & .Does the audit trail identify the authority or delegated authority identified in legislation? & Canada \\
    & & independent audit & .Do you ensure that the AI system can be audited independently? & EU \\   
    & redress & redress mechanism & .Do you establish an adequate set of mechanisms that allow for redress in case of the occurrence of any harm or adverse impact? & EU \\
    &  &  & .Do you put mechanisms in place both to provide information to (end-)users/third parties about opportunities for redress? & EU \\
    & trade-offs & interests and values & .Do you establish a mechanism to identify relevant interests and values implicated by the AI system and potential trade-offs between them? & EU \\
    & & decision/documentation & .How do you decide on such trade-offs? Do you ensure that the trade-off decision is documented? & EU \\
    & & & .What is the extent of the risk associated with no or low documentation of performance targets or trade-offs? & NIST \\
    Transparency/explainability & traceability & traceability mechanism & .Do you establish measures that can ensure traceability? & EU \\
    & & & .What is the extent of the risk associated with incomplete documentation of AI system design, or implementation, or operation? & NSW \\
    & explainability & system explainability & .Can the AI system produce outcomes that all users can understand? & EU \\
    & & & .Is the system able to produce reasons for its decisions or recommendations when required? & Canada \\
    & & & .Do you explain, validate, and document the AI model and the output data interpreted within its context to inform responsible use and governance? & NIST \\
    & & assess explainability & .Do you assess to what extent the decisions and hence the outcome made by the AI system can be understood? & EU \\
    & communication & user communication & .Do you establish mechanisms to inform users on the reasons and criteria behind the AI system’s outcomes? & EU \\
    & & & .Do you consider communication and transparency towards other audiences, third parties, or the general public? & EU \\
    Reliability/safety & resilience & mechanism & .Do you have organizational policies and practices to foster a critical thinking and safety-first mindset in the design, development, and deployment, & EU \\
    & & & and uses of AI systems to minimize potential negative impacts? \\
    & accuracy & data quality & .Do you put in place measures to ensure that the data used is comprehensive and up to date? & EU \\
    & & & .Do you put in place measures in place to assess whether there is a need for additional data, for example to improve accuracy or to eliminate bias? & EU \\
    & adverse impact & impact of uses & .Are there any uses of the system subject to a legal or internal policy restriction? & MS \\
    & & limitations & .Are there conditions in the deployment environment that would affect the system’s performance? & MS \\
    \bottomrule
  \end{tabularx}
\end{table}

\begin{figure*}[]
  \centering
  \includegraphics[width=\textwidth]{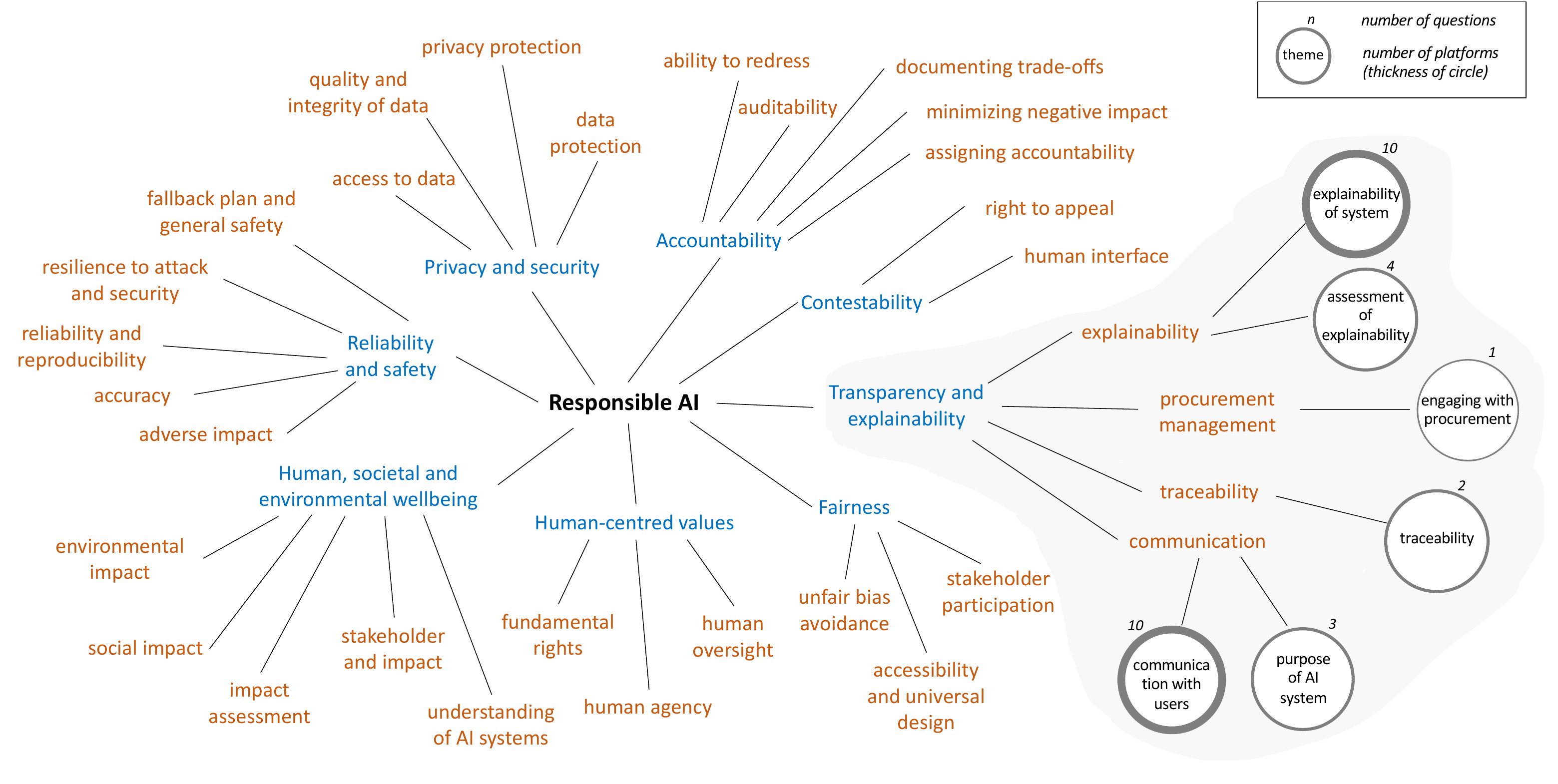}
  \caption{Risk concept map for responsible AI.}
  \label{fig:QB_Concept}
\end{figure*}

In Figure \ref{fig:QB_Concept}, the concept map represents the hierarchical organization of 31 risk themes and their distribution across eight AI principles.
Each principle consists of two to five concepts that represent crucial areas of risk requiring responsible assessment and management in AI.
%For example, the \textit{Privacy and Security} principle comprises four concepts, namely \textit{access to control, quality and integrity of data, data protection,} and \textit{privacy protection}. 
This indicates that these risk areas should be ensured and controlled throughout the entire life-cycle of AI systems.

% show the next level of concept map: e.g., Privacy or..
The expanded concept map (grey area in Figure \ref{fig:QB_Concept}) delves into 68 sub-themes within each primary risk category for in-depth exploration and understanding of specific risks.
%This allows for a more in-depth exploration and understanding of specific risks associated with RAI in each area.
%In Figure \ref{fig:QB_Concept} the grey area demonstrates the expanded concept map for \textit{Transparency and Explainability}. 
Circle shapes represent the sub-themes and indicate the number of associated risk questions.  
The thickness of each circle represents the number of source frameworks associated with each theme. 
Thicker circles suggest common coverage across frameworks and highlight areas that warrant more attention in the risk assessment process.
Similarly, circles with a higher number of questions indicate that these themes have more potential risks associated with them, underscoring their importance in risk management. This approach supports evidence-based decision-making to prioritize risk areas.

\begin{figure*}
\centering
  \includegraphics[width=\textwidth]{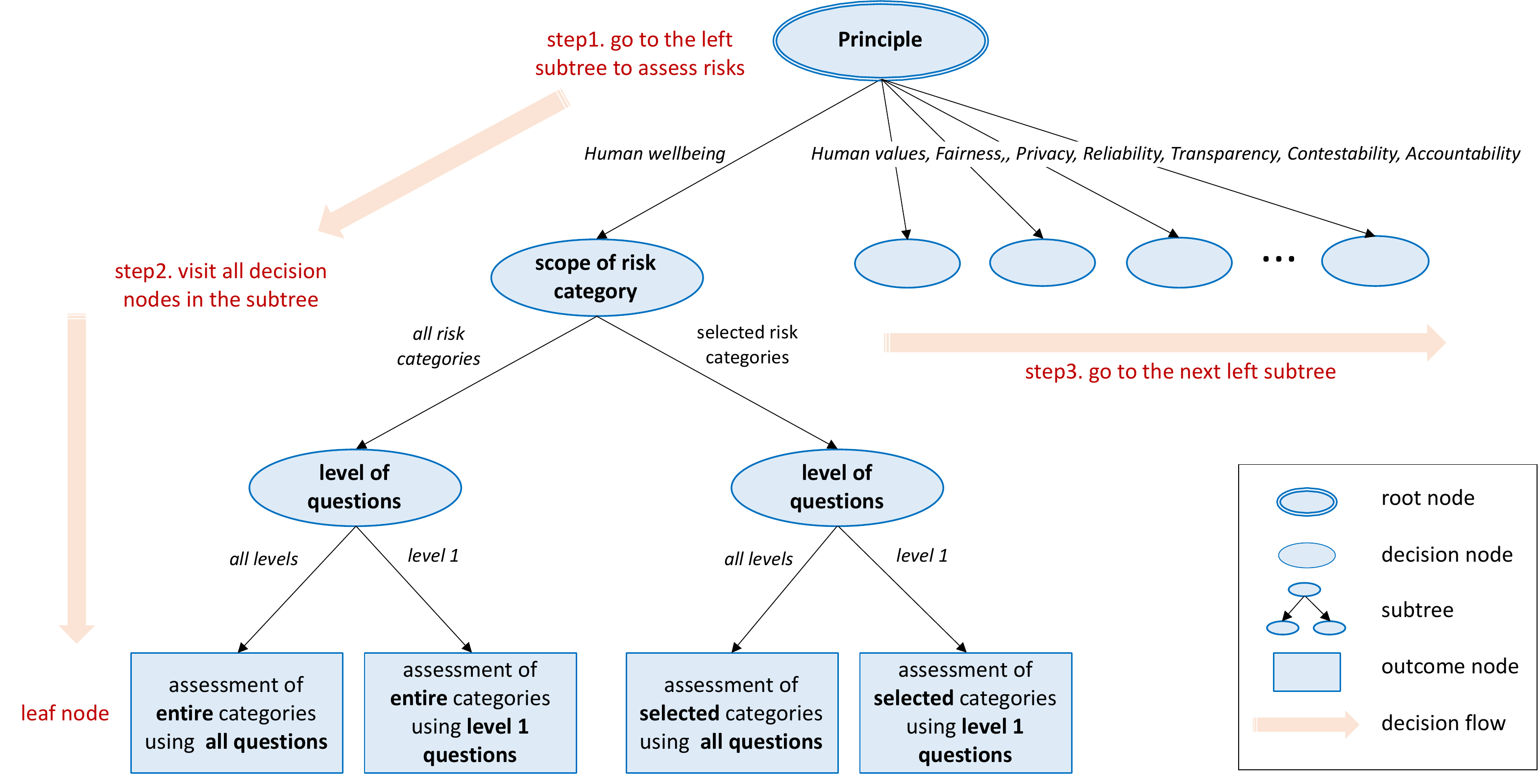}
  \caption{Overview of the decision tree for AI risk assessment.}
  \label{fig:QB_Tree}
\end{figure*}

\subsubsection{Tiered risk assessment} \label{sec:tiered}
The notion of tiered risk assessment arose from the interest in examining risks at different levels of detail while effectively utilizing the large number of questions available, as it may not be feasible to ask all of them.
% Existing frameworks may not need to consider this as much, since they have a smaller set of questions.
Existing frameworks may not prioritize this aspect as much since they have a smaller set of questions. 
%The question banks is designed to assess risks at different levels of detail or complexity.
The level of questions has been determined based on their relative importance and specificity within the category.
This accommodates the needs of different groups of stakeholders in the risk assessment process and provides a top-down approach that allows organizations to focus on the most important and high-level risks first.

The top-level question in each category is regarded as the most important and fundamental question that provides a broad overview of the key risks.
%and considerations associated with the risk category.
These questions are primarily intended for high-level decision-makers such as C-levels.

Subsequent questions are organized into second and third levels.
Second-level questions are typically more specific than the top-level question and help to provide additional detail and context.
They target senior managers and individuals with specific roles in overseeing AI-related projects.
Third-level questions are even more detailed and represent sub-questions that help to further break down the key risks.
These questions are aimed at practitioners who require a detailed understanding of the technical and operational aspects of AI risk assessment.

For example, one of the top-level questions for Transparency and Explainability, \textbf{”Does the outcome result in something that all users can understand?”}, is an import and high-level question. However, it does not delve into specific considerations and risks in the system’s development.
An possible second-level question is \textbf{"Do you design the AI system with interpretability in mind from the start?"}. 
This question assesses whether the project team considers "transparency by design" and manages potential design issues in the early development stages. 
As such, this question is suitable for the management-level stakeholders rather than C-levels.
\textbf{”Do you research and try to use the simplest and most interpretable model possible for the AI system?”} can be used to further explore the topic if the development team has conducted sufficient research on interpretability during the model selection process.

\subsubsection{Streamlined risk assessment}

With a large number of questions, QB4AIRA is essential to streamline the risk assessment process and enhance efficiency while maintaining its effectiveness in identifying and addressing potential risks \cite{baquero2020derisking}.
To achieve this goal, we have developed a hierarchical structure using a decision tree, which effectively and efficiently selects the most relevant risk questions from the extensive pool\cite{song2015decision}. 

Figure \ref{fig:QB_Tree} depicts the process of exploring risk questions (step 1-3) based on a preorder traversal approach \cite{adams1973another}. 
This initiates by visiting the root node, enabling the determination of which principles should be assessed. 
%This decision is based on the contextual factors associated with the AI systems.
It then moves to the first left subtree before proceeding to the right subtrees.
Within each subtree, based on the scope of the risk category (the first decision node of the subtree being visited), the top-level question takes precedence (level 1). 
Subsequently, a decision can be made to either proceed to the next category or visit all nodes within the current category. 
The leaf nodes of the decision tree indicate whether the risk assessment should be performed solely on the selected categories or on all categories encompassed in the tree, depending on the decision made at the preceding nodes.

\section{Evaluation}

\subsection{Case study 1: AI system risk assessment}
\subsubsection{Description}
We have assessed AI ethical risks for eight AI projects selected within a national research organization: PR1 to PR8. 
These projects primarily focus on research and include AI as a component within a larger research framework. 
The projects have different objectives; for example, PR1 aims to provide a comprehensive design tool for AI and PR3 uses AI for drug discovery. 
%It is important to note that the successful commercialization of these projects is required for the AI technology to reach the general public. 
Since the projects are currently in their early stages, they present an ideal opportunity to incorporate best practices in RAI for future development.

\subsubsection{Risk assessment}
Prior to conducting the risk assessment, we developed a dedicated risk register template that includes essential components such as risk ID, risk category, risk title, description, risk causes, and interview questions which were primarily derived from QB4AIRA.
The risk assessment sessions were scheduled between January and February 2023. Each session involved an interview lasting approximately 1.5 hours. 
The participants included a risk assessor, 1-2 observers, and 1-4 interviewees.
The interview results were analyzed to identify and assess the risks associated with AI ethics. 
%The risks were scored using the enterprise risk framework.
We have found that specific projects (e.g., PR4 and PR8) exhibit high risks related to privacy, reliability, and transparency, while the majority of projects have low risks in terms of contestability.

\subsubsection{Feedback for improvement}
Following the risk assessment process, valuable feedback was received from the stakeholders. 
It was acknowledged that while the questions in QB4AIRA proved useful in identifying risks, some questions were not directly relevant to the current stage of the projects. 
As a result, the stakeholders recommended an improvement in QB4AIRA's structure by incorporating a dimension of development stage. 
%This enhancement would enable a more efficient and effective risk identification process, aligning the questions with the specific stage of each project.
%The suggestion to introduce a development stage dimension within the question bank holds significant promise for refining the risk assessment process. 
By tailoring the questions to the specific stage of each project, potential risks can be identified more accurately and in a manner that is more applicable to the project's context. 
%This would not only enhance the relevance and usefulness of QB4AIRA but also optimize the allocation of resources and efforts in addressing AI ethical risks.

%By heeding this valuable feedback and integrating a development stage dimension into the question bank, the risk assessment process can be further optimized to address the specific risks and challenges that arise at different stages of AI project development. 
This enhancement would improve the applicability, efficiency, and effectiveness of QB4AIRA, contributing to more robust and comprehensive AI risk management practices.

% Case 2. Smart risk assessment
%    - Explain how our question bank serves as a knowledge base to support smart/automated risk assessment.
%    - Explain a breif process/use cases for it; e.g., building a knowledge graph for training the local model and/or selecting scenario-based questions and provide an efficient/effective approach to self-risk assessment
%   - Show what feedbacks/improvements have been recieved: e.g., need to consider metrics/scoring mechanisms, etc.
\subsection{Case study 2: Smart risk assessment tool}
\subsubsection{Description}
The Smart Risk Assessment Tool (SRA) is an innovative chatbot designed to provide users with answers to their questions regarding AI risk assessment. 
The development of SRA draws upon the analysis of several hundreds of AI incident cases, which have been collected and analyzed to construct a comprehensive knowledge graph.
The current version of SRA serves as a prototype, demonstrating its capabilities and potential as a proof of concept. 
It utilizes ChatGPT as a foundation model to generate responses and future iterations will involve the utilization of fine-tuned local models, enhancing the accuracy and contextual relevance of the tool.

\subsubsection{Smart risk assessment}
SRA is currently using a selection of questions from QB4AIRA to demonstrate its functionality to stakeholders, showcasing the potential of integrating the comprehensive question bank into the tool's self-risk assessment capabilities.
Through a meeting with the development team, we discussed 
%the integration of QB4AIRA and SRA to ultimately facilitate \textit{smart self-risk assessment}. 
how to leverage QB4AIRA as an input to build a knowledge graph within the tool.
It include i) utilizing \textit{tiered risk assessment} to allow the tool to adapt and respond to users' specific contexts, %such as the type of AI, current development stage, and stakeholder involved
ii) \textit{decision tree functionality} to implement streamlined risk assessment to intelligently select a series of questions based on the user's context and requirements, and iii) \textit{risk metrics} to quantify and evaluate the identified risks.

\subsubsection{Feedback for improvement}
QB4AIRA currently lacks concrete risk metrics to support SRA and potential external applications. We acknowledge this valuable feedback as an important input for the future evolution of QB4AIRA, aiming to enhance its usability and effectiveness in measuring and mitigating AI-related risks.

% Need to consider Case 3.
% Case 3. Educational card game
%   - Explan how the question bank can be used for the card game: e.g., provide additional design features by identifying potential risks and vulnerabilities.
% ???? at different stages of development.

% Need to consider the following cases as well.
% Case 4. ESG risk assessment
% Case 5. BigCode project: for specific type of AI project (Open community and LLM)

% Sunny- this section should be rewritten.
\section{Conclusion}

%In this study, we introduced QB4AIRA, a comprehensive and systematic question bank for conducting AI risk assessments. 
QB4AIRA is developed via collecting and refining questions from five extant AI risk frameworks, it includes 293 questions that cover a wide range of AI risk areas and can be used by different stakeholders, including AI developers, policymakers, and regulators. 
To prioritize and align the questions in QB4AIRA with the context of AI systems, we assessed and determined their level of importance and details. This process aimed to ensure efficient and effective risk assessment while addressing specific needs.
By providing a standardized approach to AI risk assessment, QB4AIRA can help ensure that AI systems are developed and deployed in a responsible and ethical manner that addresses potential risks and harms. 
Furthermore, it can serve as a valuable resource for the development of new AI risk frameworks and guidelines in the future.

In future work, we plan to tailor questions to the specific stage to effectively address and mitigate potential risks. 
Further, we will propose appropriate risk assessment metrics and methods for scoring risks and avoiding subjective decision-making. 
We would like to emphasize that QB4AIRA is designed to be adaptive and expandable. 
Our future plans include incorporating additional AI risk frameworks,  regulations and standards such as \textit{EU AI Act \footnote{https://artificialintelligenceact.eu/} and ISO for AI risk} \footnote{https://www.iso.org/standard/77304.html} to further enhance the coverage and relevance of the question bank.

%Bibliography
\bibliographystyle{unsrt}  
\bibliography{main}

\end{document}